\documentclass[9pt,twocolumn]{article}
\pdfoutput=1
\usepackage[T1]{fontenc}
\usepackage{fancyhdr}
\usepackage[colorlinks=true,linkcolor=red,citecolor=blue,urlcolor=blue]{hyperref}
\usepackage{caption} 
\usepackage{color} 
\usepackage{verbatim}
\usepackage{graphicx}
\usepackage{geometry}
\usepackage{booktabs}
\usepackage[sort&compress,super]{natbib}
\title{Proper network randomization is key to assessing social balance}

\author{Bingjie Hao$^1$ \and István A. Kovács$^{1,2,*}$ }
\date{
{\footnotesize
$^1$Department of Physics and Astronomy, Northwestern University, Evanston, IL 60208\\
$^2$Northwestern Institute on Complex Systems, Northwestern University, Evanston, IL 60208\\
$^*$\href{mailto:istvan.kovacs@northwestern.edu}{istvan.kovacs@northwestern.edu}}
}

\begin{document}

\twocolumn[
  \begin{@twocolumnfalse}
    \maketitle
    \begin{abstract}
    Studying significant network patterns, known as graphlets (or motifs), has been a popular approach to understand the underlying organizing principles of complex networks. Statistical significance is routinely assessed by comparing to null models that randomize the connections while preserving some key aspects of the data. However, in signed networks, capturing both positive (friendly) and negative (hostile) relations, 
    the results have been controversial and also at odds with the classical theory of structural balance. We show that this is largely due to the fact that large-scale signed networks exhibit a poor correlation between the number of positive and negative ties of each node. 
     As a solution, here we propose a null model based on the maximum entropy framework that preserves both the signed degrees and the network topology (STP randomization). With STP randomization the results change qualitatively and most social networks consistently satisfy strong structural balance, both at the level of triangles and larger graphlets. We propose a potential underlying mechanism of the observed patterns in signed social networks and outline further applications of STP randomization.
    \end{abstract}
  \end{@twocolumnfalse}
]

\section*{Introduction}
Individuals within society can be viewed as nodes in a social network, with edges representing various relationships between them. These relationships are highly diverse in nature and can be often expressed in either positive (friend/trust) or negative (foe/distrust) terms \cite{leskovecSignedNetworksSocial2010}, leading to signed social networks, with varying degrees of polarization \cite{huangPOLEPolarizedEmbedding2022}. 
Quantifying the wiring patterns is the first step towards understanding why certain connections are formed and not others, i.e., the underlying wiring mechanisms, as well as towards understanding and potentially reducing polarization in social media \cite{huangPOLEPolarizedEmbedding2022, garimellaLongTermAnalysisPolarization2017,gillaniMeMyEcho2018,tuckerSocialMediaPolitical2018,grossDenseNetworkMotifs2023}.
%
As a key concept, network graphlets (and motifs) \cite{miloNetworkMotifsSimple2002,ahmedGraphletDecompositionFramework2017} are fundamental patterns of connections that occur significantly  more frequently than in a \emph{null model}, which is a suitably randomized version of the empirical data \cite{miloNetworkMotifsSimple2002}. 
Graphlets, also known as induced subgraphs \cite{przuljBiologicalNetworkComparison2007}, specify the existence and sign of each link within a subset of nodes. In contrast, motifs (or non-induced subgraphs) \cite{miloNetworkMotifsSimple2002}, specify only the required links, allowing for the presence or absence of other links.
For instance, in an undirected signed network, a graphlet consisting of three nodes connected by two edges indicates the absence of the third link, while a motif encompasses instances with or without the third link. Note that any fully connected graphlet can also be referred to as a motif.

Seminal studies have shown that network graphlets and motifs play an important role in understanding the organization, functionality, and hidden mechanisms behind many complex systems, from social networks to brain connectivity and protein-protein interaction networks \cite{miloNetworkMotifsSimple2002,shen-orrNetworkMotifsTranscriptional2002,tranCountingMotifsHuman2013,liInhomogeneousHypergraphClustering2017,chenNeMoFinderDissectingGenomewide2006,spornsMotifsBrainNetworks2004, ahmedGraphletDecompositionFramework2017, przuljBiologicalNetworkComparison2007}.

Fully connected `triangle' graphlets of three nodes are particularly informative on tie formation mechanisms between acquaintances of the same node.  
As a starting point, strong structural balance (SB) \cite{cartwrightStructuralBalanceGeneralization1956} captures the intuitive notions of ``the friend of my friend is my friend", ``the enemy of my friend is my enemy'', and ``the enemy of my enemy is my friend''. All these examples correspond to \emph{balanced} cycles (a path that starts and ends at the same node, without revisiting any other nodes in between) of length three, where the product of edge signs along the cycle in the network is positive. This expectation has been extended to cycles of any length, stating that a network is maximally balanced if all cycles are balanced \cite{cartwrightStructuralBalanceGeneralization1956}, including cycles of length four, corresponding to `square' graphlets. In practice, there are often deviations from maximal balance, requiring the statistical analyses of enrichment of the studied patterns versus a null model. 
Although it is generally believed that social networks tend to be in somewhat balanced states \cite{situngkirSocialBalanceTheory2004}, the conclusions strongly vary depending on the studied datasets and the chosen null model \cite{szellMultirelationalOrganizationLargescale2010,raoMarkovChainMonte1996,singhMeasuringBalanceSigned2017,fengTestingBalanceSocial2022}.

\begin{figure}[htpb]
    \centering
    \includegraphics[width=\linewidth]{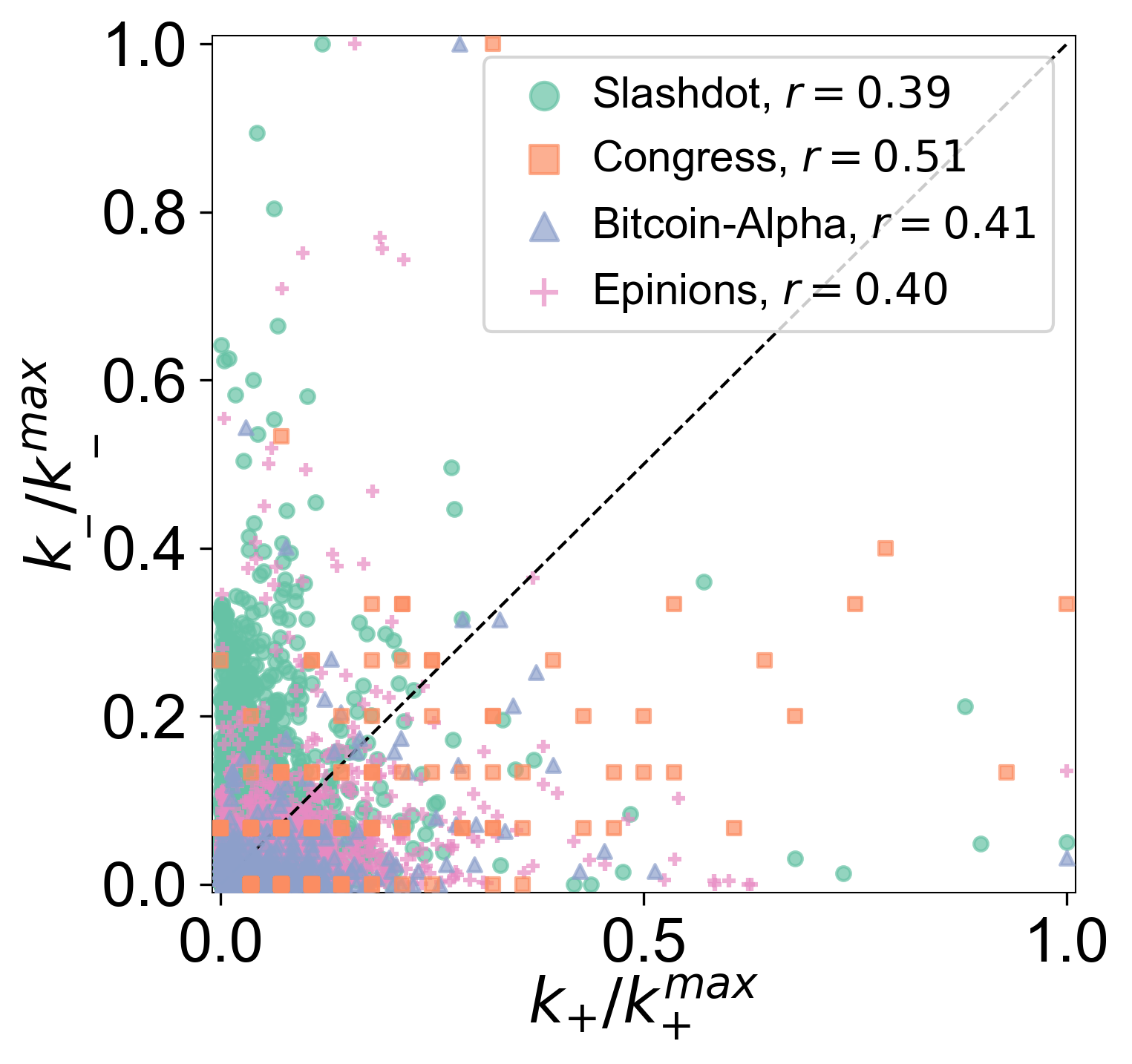}
    \caption{Signed degree inconsistency. Positive ($k_{+}$) and negative degrees ($k_{-}$) in social networks are poorly correlated. The dashed black line indicates a perfect correlation between $k_{+}$ and $k_{-}$. The $r$ values denote the Pearson correlation coefficient between $k_{+}$ and $k_{-}$ of each dataset.}
    \label{fig:intro}
\end{figure}

As a basic example, the `rewire' null model \cite{raoMarkovChainMonte1996} swaps edges between nodes while preserving the node degree ($k$, number of neighbors), leading to networks with disrupted topology. 
Hence, the conclusions based on the rewire null model mix the pattern formation mechanisms arising from edge signs with those of purely topological origin. 
A more commonly used approach is the `sign shuffle' null model \cite{szellMultirelationalOrganizationLargescale2010}. In this null model, the topology of the network and the total number of positive and negative edges are preserved, while the sign of each edge is randomly assigned according to the observed fraction of positive edges in the input network. 
Note that this method has the limitation that 
all nodes are assumed to have the same ratio of positive edges. 
As illustrated in Fig.~\ref{fig:intro}, this assumption is far from reality. Indeed, in real-life networks, some nodes are more `friendly' than others, i.e., holding mostly positive edges. 

In this paper, we show that the current discrepancies arise from the fact that an adequate null model needs to preserve both the network topology and the expected signed degree of each node, while state-of-the-art null models can only preserve one of these constraints.
As a solution, we propose an alternative network ensemble,  
a Signed degree and Topology Preserving (STP) null model based on the maximum entropy framework (see Methods), that simultaneously preserves the network topology as well as the mean positive and negative node degrees.
%

We examine the impact of network randomization on a collection of signed social networks covering datasets of various scales, including (i) Slashdot - a friend/foe network in the technological news site Slashdot \cite{leskovecSignedNetworksSocial2010}; (ii) Congress - a political network where signed edges represent (un/)favorable interactions between U.S. congresspeople on the House floor in 2005 \cite{huangPOLEPolarizedEmbedding2022}; (iii) Bitcoin-Alpha - a trust/distrust network of Bitcoin traders on the platform Bitcoin Alpha \cite{kumarEdgeWeightPrediction2016}; and (iv) Epinions: a trust/distrust network among users of the product review site Epinions \cite{leskovecSignedNetworksSocial2010}. For an overview of the fundamental properties of these datasets, see Table \ref{tab:overview}. As a key observation, positive ($k_{+}$) and negative ($k_{-}$) node degrees have a low correlation in all four networks (Fig.~\ref{fig:intro}), indicating that null models that do not consider the signed degree as a confounding factor may lead to biased results. 
On these examples, we 
show that the STP null model changes the results qualitatively, leading to a consistent interpretation of patterns in signed social networks. 
We conclude by discussing potential underlying pattern formation mechanisms behind our observations, as well as further applications and extensions of STP randomization.

\section*{Results}


 %

\begin{table*}[htbp]
\centering
\caption{Basic characteristics of the studied signed network datasets}
\begin{tabular}{lrrrrrr}
    \toprule
    Dataset & Slashdot & Congress & Bitcoin-Alpha & Epinions &      SB Reference & EC\\
    \midrule
    Nodes          &    $82\,052$ &      $219$ &          $3\,766$ &   $119\,070$ &    $3\,000$ &    $3\,000$ \\
    Edges          &   $498\,527$ &      $520$ &         $13\,872$ &   $701\,569$ &    $9\,543$ &    $7\,042$ \\
    Density        &   $0.0001$ &   $0.0218$ &         $0.002$ &   $0.0001$ &  $0.0021$ & $0.0015$ \\
    Positive ratio &   $0.764$ &   $0.796$ &         $0.917$ &   $0.832$ &  $0.754$ & $0.753$\\
    \bottomrule
\end{tabular}

\label{tab:overview}
\end{table*}

\begin{figure*}[t]
    \centering
    \includegraphics[width=\linewidth]{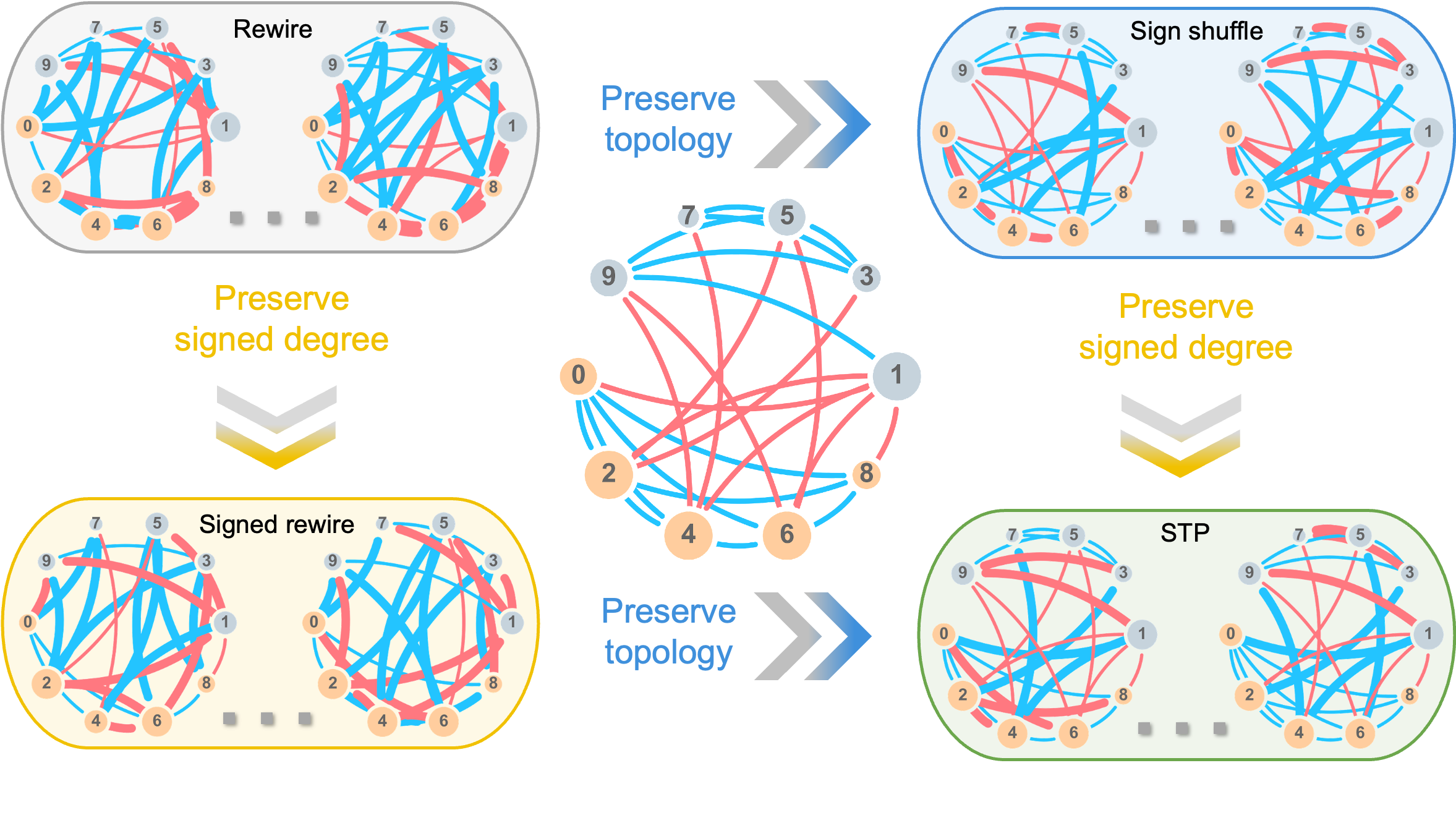}
    \caption{Overview of signed null models. The original network that contains two groups of nodes (yellow and grey) is shown in the middle. Positive edges are shown in blue, while negatives in red. 
    Thicker lines indicate edges that are different from the original network.
    }
    \label{fig:null-models}
\end{figure*}

\subsection*{Signed null models}
To investigate how the topology and signed degree affect the graphlet statistics, we consider four null models for signed networks, see Fig.~\ref{fig:null-models}. In addition to the commonly used rewire and sign shuffle null models and our STP null model, we also consider the `signed rewire' null model \cite{liConstructingRefinedNull2021}. 
The signed rewire null model preserves the signed degree of each node by rewiring the positive and negative subgraphs separately. As a result, the topology is not preserved and there can be edges in the signed rewire null model that are both negative and positive (see Methods). In Fig.~\ref{fig:null-models}, we illustrate the studied null models on a toy network satisfying SB. This toy network
contains two groups of nodes (indicated by different node colors), with positive edges among group members and negative edges between the groups \cite{kirkleyBalanceSignedNetworks2019}. Note that some nodes are more friendly (like node 0) or more aggressive (like node 1), i.e., have a higher fraction of positive or negative edges than others. 

\subsection*{Signed triangle patterns}
\begin{figure*}[htpb]
    \centering
    \includegraphics[width=\linewidth]{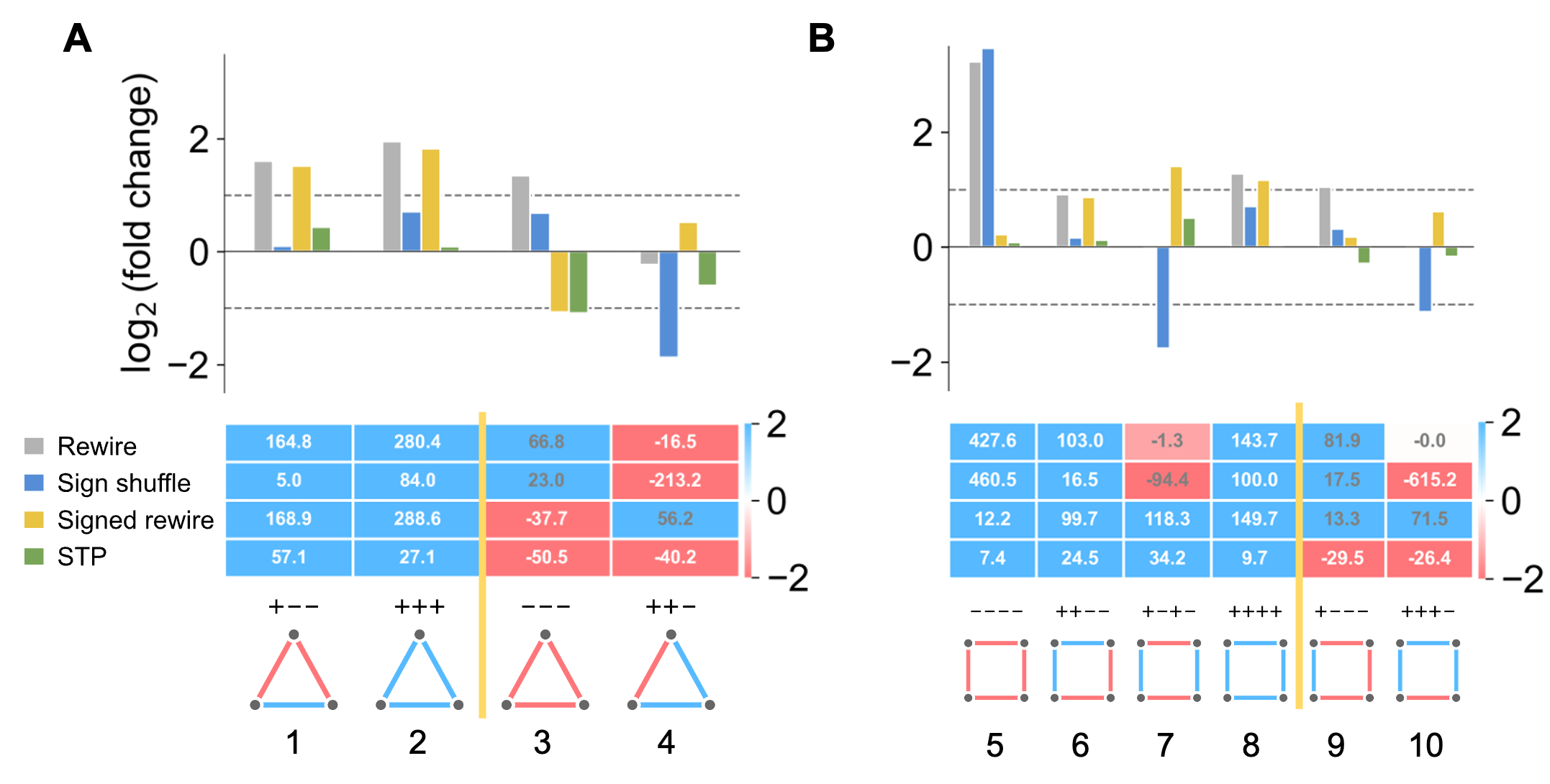}
    \caption{Signed graphlets in the Slashdot network compared to different null models. (A) Triangles. (B) Squares. The $\log_2(\mathrm{fold~change})$ is shown on the top accompanied by the grey dashed line indicating a 2-fold increase or decrease. $z$-scores are shown at the bottom, in white if matching SB expectations, and black otherwise. The background of the $z$-scores is blue for positive values and red for negative values.}
    \label{fig:tri_sq}
\end{figure*}

\begin{figure*}[htpb]
    \centering
    \includegraphics[width=0.6\linewidth]{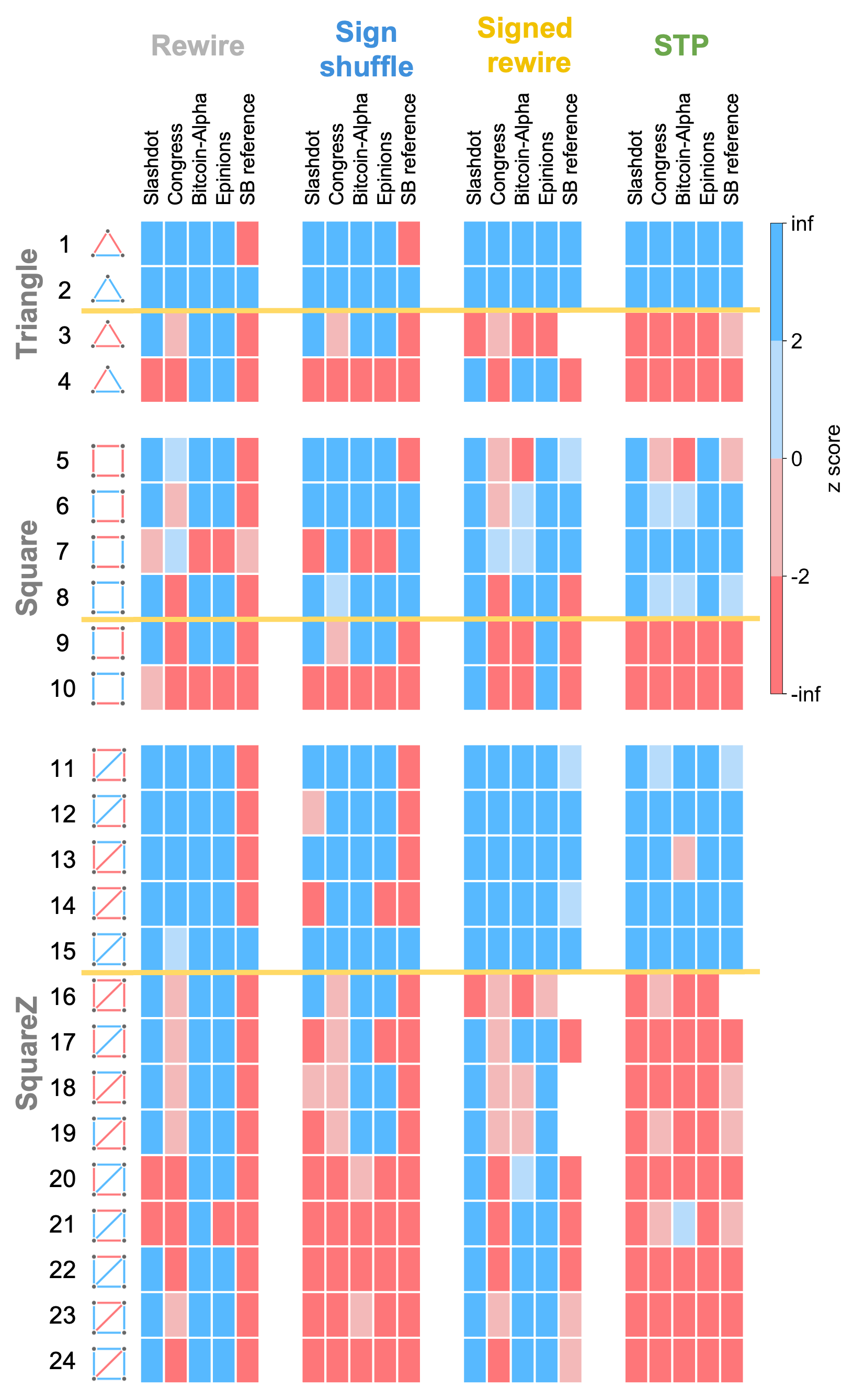}
    \caption{Overview of graphlet significance in the studied networks. The $z$-scores are indicated by blue (overrepresented) and red (underrepresented) blocks. We list the balanced graphlets first, separated from the unbalanced graphlets by a yellow line. We leave the block white if $n_{obs} = \sigma_{rand} =0$ as it leads to an undetermined $z$-score.}
    \label{fig:result_summary}
\end{figure*}

To test social balance in real networks, we first consider signed triangles as illustrated in Fig.~\ref{fig:tri_sq}A for the Slashdot network. Each triangle graphlet (or equivalently motif) is counted and compared to the expected number of such triangles in four different null models. The fold change is calculated as
\begin{equation}
    \mathrm{fold~change} = n_{obs} / <n_{rand}>\;,
\end{equation}
indicating the relative graphlet frequency ($n_{obs}$) in the original network compared to the frequency in each randomized version ($<n_{rand}>$) in Fig.~\ref{fig:tri_sq}. As a standard measure of statistical significance, we calculate the $z$-score as
\begin{equation}
    z = \frac{n_{obs}-<n_{rand}>}{\sigma_{rand}}\;,
\end{equation}
where $\sigma_{rand}$ denotes the standard deviation of $n_{rand}$. 
In the Slashdot network, both the rewire and sign shuffle null models would claim that only $++-$ triangles are underrepresented. This conclusion aligns with the notion of weak structural balance (WB) \cite{davisClusteringStructuralBalance1967,leskovecSignedNetworksSocial2010,esmailianMesoscopicAnalysisOnline2014}, which alleviates the structural balance so that triangles with exactly one negative edge should be underrepresented. In other words,  WB relaxes the assumption that ``the enemy of my enemy is my friend''.
On the contrary, the signed rewire model would conclude that only $---$ triangles are underrepresented. 
%
%
At the same time, the STP null model identifies that both $---$ and $+--$ triangles are heavily underrepresented, in line with SB. 

The same conclusion is observed consistently in the other studied networks as well, when both the signed degrees and the topology is conserved, meaning that 
signed social networks are strongly balanced at the triangle level according to STP randomization (Fig.~\ref{fig:result_summary}). 
No consistent conclusions can be drawn from the rewire and signed rewire null models, most likely due to the fact that the topology is disrupted in these cases.
The sign shuffle null model appears to be consistent with WB in all real networks.  
However, as a clear shortcoming, the sign shuffle and rewire null models fail to detect SB in our model network explicitly built as an SB reference \cite{hararyMeasurementStructuralBalance1959} (see Methods). 
To sum, the known null models lead to conflicting and inadequate conclusions at the triangle level. This incoherence is further amplified when analyzing four-node graphlets, as discussed next.


\begin{figure*}[htpb]
    \centering
    \includegraphics[width=0.8\linewidth]{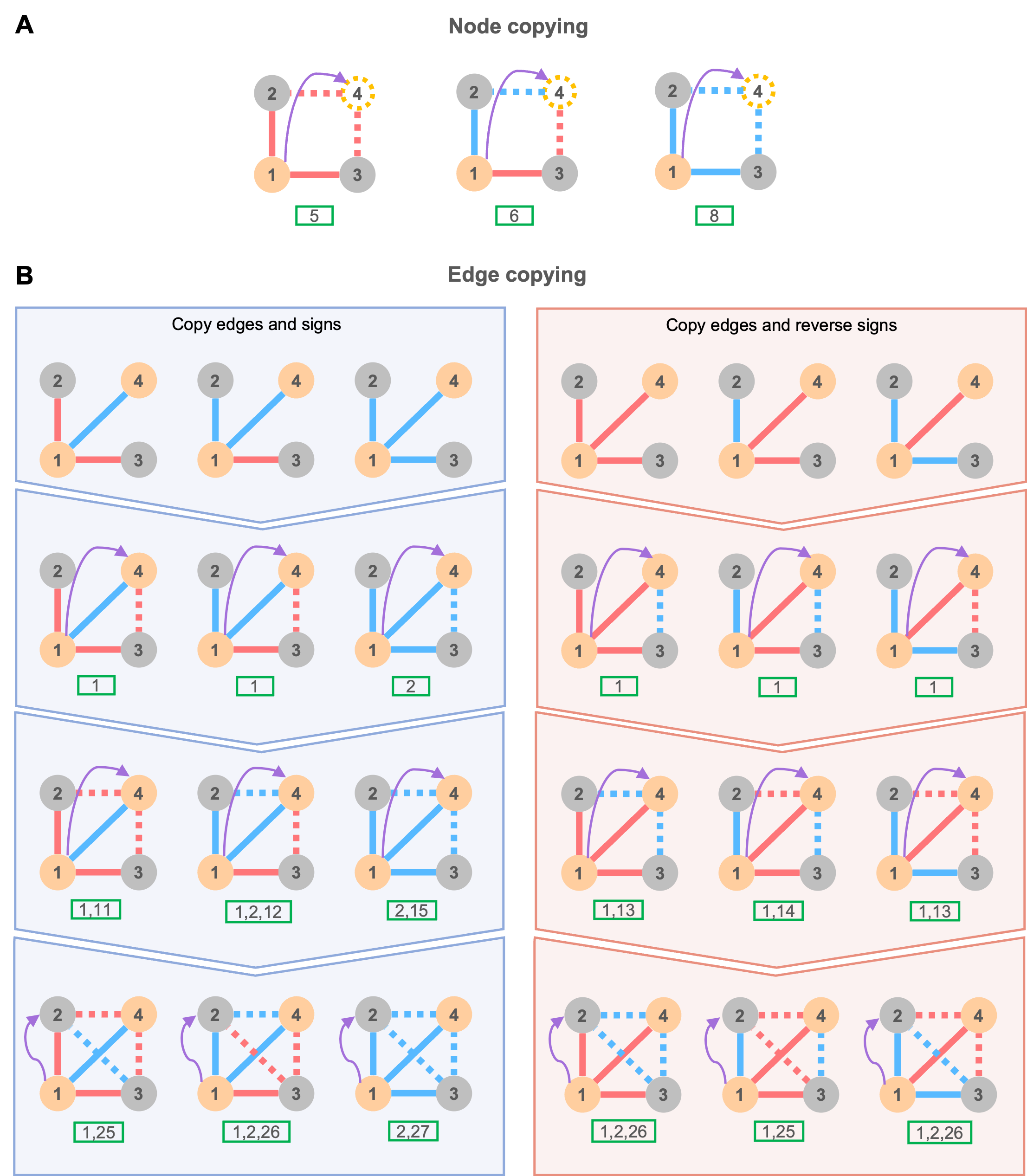}
    \caption{Illustration of signed copying mechanisms. (A) Signed node copying. When node 4 is added to the existing network, it copies the edges and signs of node 1, forming squares. Note that graphlet 7 ($+-+-{}$) can not be generated this way. (B) Signed edge copying. When node 4 is connected to node 1 by a positive edge, it may copy node 1's edges and signs, forming triangles, squareZs and squareXs. When node 4 is connected to node 1 by a negative edge, it may copy node 1's edges but will reverse the signs. The edge copying mechanism forms balanced triangles, eventually leading to larger balanced graphlets. The initial edges are indicated by solid lines and the copied edges are indicated by dotted lines. The resulting graphlets are indicated by the indices within the green boxes, following the notation of Fig.~\ref{fig:result_summary} and Fig.~S1. The purple arrows point from the node being copied to the node that is copying.}
    \label{fig:square_mechanisms}
\end{figure*}

\subsection*{Signed square patterns}

We start the analyses of four-node graphlets by investigating six cases of square graphlets 
(Fig.~\ref{fig:tri_sq}B). In the Slashdot network, STP randomization again indicates consistency with SB, while the conclusions of other null models would differ qualitatively. However, understanding SB at the square level requires further efforts.
Squares are traditionally less investigated than triangles for multiple reasons. In addition to the increased computational complexity, it appears less realistic to assume that individuals know their social networks at the square level \cite{kirkleyBalanceSignedNetworks2019}. Furthermore, comparing the observed frequencies of signed squares to the current null models often leads to inconsistent conclusions (Fig.~\ref{fig:result_summary}). 
Indeed, all square graphlets can be either significantly over- or underrepresented depending on the choice of network data and the null model. Existing null models again fail to detect SB at the square level even in the SB reference network. 
In contrast, when compared to the STP null model, all unbalanced squares are consistently underrepresented, while most balanced squares, with the exception of the $----{}$ graphlet, are overrepresented. 
While the $----{}$ graphlet appears to be underrepresented, it is just barely so at these network sizes, as even the strongest signal in Bitcoin-Alpha is just $z\sim2$. 
%
%
Overall, the square graphlet results with STP randomization indicate that social networks are compatible with SB, with the potential exception of the $----{}$ graphlet. 


\subsection*{Potential mechanisms behind signed patterns}

Next, we show that the consistent results obtained by using the STP null model enable the formulation of hypotheses for potential wiring mechanisms.
Previous studies attempted to explain square graphlet statistics by a node copying mechanism \cite{bhatDensificationStructuralTransitions2016}. We first generalize the node copying mechanism to signed networks, where a new node can replicate both the connections of existing nodes in the network and the corresponding edge signs. This mechanism contributes to the formation of balanced squares, as illustrated in Fig.~\ref{fig:square_mechanisms}A. However, our results in Fig.~\ref{fig:result_summary} show that the square $+-+-{}$ graphlet is also overrepresented in all studied networks, suggesting that this simple node copying mechanism is not sufficient to explain the full extent of our observations. Moreover, signed triangle graphlets are not necessarily explained by a node copying mechanism, expected to depend also on the initial conditions. 

Thus, as a potential solution, we propose an edge copying mechanism as shown in Fig.~\ref{fig:square_mechanisms}B. As a basic step, a node can copy one of its friend's connections. In other words, nodes connected by positive edges are assumed to copy each other's attitude towards other nodes. Conversely, negatively linked nodes replicate the edges of their foes while reversing the sign. 
The process of edge copying initially results in the formation of balanced triangles and eventually leads to larger balanced graphlets, such as `squareZ' graphlets, which refer to squares with an additional edge. 
Eventually, this process leads to the formation of fully connected four-node subgraphs, referred to as `squareX' graphlets/motifs, when other nodes also copy edges in the same subnetwork. 

The proposed edge copying mechanism is compatible with the squareZ graphlet results shown in Fig.~\ref{fig:result_summary}. 
Comparing the observed squareZ statistics to the STP null model, all datasets consistently agree with the SB expectation, while there is no consistent conclusion from other null models (Fig.~\ref{fig:result_summary}). In addition to squareZ, we have also considered the fully connected four-node squareX graphlet/motifs as shown in Fig.~S1.
When compared to the STP null model, balanced (unbalanced) squareXs are again overrepresented (underrepresented), a pattern that is not observed with other null models.
We also implemented the edge copying mechanism to generate an edge copying reference network, EC (see Methods). Note that with the chosen initial conditions, EC also satisfies SB. 
Yet, in general, the edge copying mechanism does not necessarily lead to the formation of balanced square graphlets. 
This finding is actually in line with our observations, 
as the statistics 
of $----{}$ appears to vary across datasets. 
Yet, our analysis indicates that $----{}$ square motifs with one or two additional positive edges are 
overrepresented, as suggested by 
the edge copying mechanism. 

The proposed edge copying mechanism on its own leaves the question of balanced squares open. To potentially elucidate the presence of mostly balanced squares in the studied datasets, we also explore the implications of edge copying between disconnected nodes. In this scenario, a node may connect to the neighbors of another (unrelated) node with or without reversing the edge signs.  
Irrespectively of the frequency of edge reversals in this step, it will lead to balanced squares of the form $++--{}$. Events with no sign reversal also contribute to $++++{}$ and $----{}$, in addition to $++--{}$.
Notably, as long as some sign reversal occurs, it leads to the enrichment of $+-+-{}$ graphlets, contrary to the above-discussed node copying mechanism. Note that this particular scenario of edge copying is more feasible when a node can access information on the signed edges of other unrelated nodes. 
Intriguingly, this can potentially explain why the square graphlet results depend on the datasets. For example, in Slashdot, one can easily know others' friends and foes without establishing a direct connection, potentially resulting in more balanced squares. Under other circumstances, individuals may have access to a strangers' friend lists but may have limited access to a strangers' blacklists (or foes) \cite{cramerLogicalApproachRestricting2015}, potentially leading to the underrepresentation of $----{}$ squares.

\section*{Discussion}
Graphlet statistics provides key insight into mechanisms of network wiring and function. However, it is important to interpret the results in the context of an adequate null model.
Up until now, signed network null models had a crucial shortcoming as they either ignored the signed degree preferences of the nodes or the network topology. In this study, we proposed the STP null model that preserves both the signed degrees and the network topology. We found that the STP null model provides more consistent results across signed social networks than previous methods, favoring SB at both the level of triangles and four-node graphlets, with the potential exception of the $----{}$ graphlets. Note that the analysis of various motifs (instead of graphlets) arrives at the same qualitative conclusions (Fig.~S3).

We also proposed a signed edge copying mechanism that results in the formation of balanced triangles, squareZ and squareX graphlets, in line with our observations.  
Note that edge copying provides a simple, yet plausible, example of forming balanced four-node patterns, questioning the current paradigm that ignores four-node mechanisms \cite{kirkleyBalanceSignedNetworks2019}.
Even so, without following the dynamics of these networks, we cannot fully conclude that edge copying is actually at play in these networks. This a point that needs to be further explored in future work.


STP randomization has widespread potential applications and extensions. To start, it can provide a more adequate alternative null model to quantify balance \cite{kirkleyBalanceSignedNetworks2019,facchettiComputingGlobalStructural2011,estradaWalkbasedMeasureBalance2014,singhMeasuringBalanceSigned2017,marvelEnergyLandscapeSocial2009} 
or measure polarization \cite{huangPOLEPolarizedEmbedding2022,garimellaLongTermAnalysisPolarization2017,gillaniMeMyEcho2018,tuckerSocialMediaPolitical2018} in social networks. STP randomization can also be used for anomaly detection \cite{kirkleyBalanceSignedNetworks2019}, as well as to infer the existence and sign of future or missing connections \cite{harrisComputationalInferenceSynaptic2022,chiangExploitingLongerCycles2011,hsiehLowRankModeling2012,wangOnlineMatrixCompletion2017},
and to perform signed graph embeddings \cite{yuanSNESignedNetwork2017,wangSignedNetworkEmbedding2017,islamSIGNetScalableEmbeddings2018,chenBridgeEnhancedSigned2018,javariROSERolebasedSigned2020}.
In addition, the STP null model can be extended to directed and weighted networks \cite{yaoHowNetworkProperties2019}, with the potential to contrast large-scale data against alternatives to SB, such as status theory \cite{leskovecSignedNetworksSocial2010}. 


\section*{Methods}
\subsection*{Signed social network datasets}
The three large signed social networks analyzed in this study were downloaded from the Stanford Network Analysis Platform (\href{http://snap.stanford.edu/}{http://snap.stanford.edu/}): (i) Bitcoin-Alpha, the trust/distrust network among people who trade Bitcoin on a platform called Bitcoin Alpha;
(ii) Slashdot, friend/foes network of the technological news site Slashdot released on February 2009; (iii) Epinions, who-trust-whom online social network of a general consumer review site Epinions. The smaller scale Congress network is from ref. \cite{huangPOLEPolarizedEmbedding2022}.
More details of the construction of the datasets can be found 
in ref.~\cite{leskovecSignedNetworksSocial2010,kumarEdgeWeightPrediction2016}. Network edges are considered to be undirected. This process leads to only a very limited number of edge sign inconsistencies. Such inconsistent edges are disregarded in our analysis, together with any self-loops \cite{facchettiComputingGlobalStructural2011}. Only the largest connected component of each network is considered.

\subsection*{Construction of SB reference network}
We construct an SB reference network according to Harary's theorem of balance \cite{hararyMeasurementStructuralBalance1959}. The 3000 nodes in the SB reference network are first divided into two equal groups. We then generate two degree sequences according to power-law degree distributions with $\mathrm{exponent}=2$ and $\mathrm{exponent}=3$ for positive and negative degree sequences, respectively. The negative degree sequence is used to generate negative edges between members of different groups and the positive degree sequence is used to generate positive edges between members of the same groups. For the positive degree sequence, we swap each degree with another randomly picked degree in the sequence with $\mathrm{probability}=0.2$ to deliberately make positive degrees and negative degrees less correlated ($r=0.56$). The resulting SB reference network has comparable density and positive edge ratios to real-life social networks as shown in Table \ref{tab:overview}.

\subsection*{Construction of the EC network}
We use the edge copying mechanism to construct EC networks. We initialized the construction process with a balanced $+++$ triangle and subsequently added nodes to the network. Each new node connects to a randomly chosen node \cite{bhatDensificationStructuralTransitions2016}, with the sign determined by a parameter $q$, which defines the probability of a positive edge. Additionally, each new node attempts to connect to the neighbors of the selected node with a probability $p$. If the new node is connected to the selected node with a positive (negative) edge, the new node copies (reverses) the sign between the randomly chosen node and its neighbor. We used the parameter values $p=0.3, q=0.85$. The positive and negative degrees follow the power-law degree distribution shown in Fig.~S2.

\subsection*{Existing Null Models}
In the rewire null model, we randomly pick two edges $A \rightarrow B$ and $C \rightarrow D$ and try to swap the edges as $A \rightarrow D $ and $C \rightarrow B$. Such an attempt is aborted when the resulting edges form self-loops ($A\rightarrow A$) or existing edges. To achieve sufficient network randomization, we perform $4E$ edge swap attempts, where $E$ represents the number of edges in the network. In the signed rewire null model, we use the same method as in the rewire null model but only swap edges if they have the same sign, thus preserving the signed degrees of each node. This may lead to multi-edges with both positive and negative signs. In such cases, we randomly assign a sign to the edge. In the sign shuffle null model, we randomly and independently assign positive or negative signs to each edge, while preserving the total number of positive and negative edges.

\subsection*{STP Null Model}
A signed network $G$, is first divided into two subnetworks, namely the positive (negative) subnetwork $G_p$ ($G_n$) that includes all the positive (negative) edges in $G$. We then randomly select a subnetwork $G_{pr}$ from $G$ while keeping the node degrees of $G_{pr}$ the same as $G_p$ on average. The remaining network is considered as the randomized negative subnetwork $G_{nr}$. The required constrained randomization is achieved using the maximum entropy approach \cite{kovacsUncoveringGeneticBlueprint2020, chatterjeeImprovingGeneralizabilityProteinligand2023}. To construct $G_{pr}$, we fix its average node degree as the original degree in $G_p$ as $<k_i>_{Gpr} = k_i(Gp)$. The resulting probability of selecting an existing edge in $G$ to be part of the subnetwork between nodes $i$ and $j$ is given by $p_{ij} = 1/(1+\alpha_i \alpha_j)$, where $\alpha_i$ are found iteratively as 
\begin{equation}
    \alpha_i^{\prime} = \frac{1}{k_i} \sum_{j,(i,j) \in G} \frac{1}{\alpha_j + 1/\alpha_i}.
\end{equation}
The initial condition is simply $\alpha_i^{(0)} \equiv 1$ and we stop the iteration when the maximum relative change of $\alpha_i$ is less than $10^{-3}$ between two consecutive iterations.


\subsection*{Data and Code Availability}
For reproducibility, we provide code and processed data at \href{https://github.com/hbj153/signed_null}{https://github.com/hbj153/signed\_null}.

\bibliography{pnas-sample}

\begin{thebibliography}{10}

\bibitem{leskovecSignedNetworksSocial2010}
Jure Leskovec, Daniel Huttenlocher, and Jon Kleinberg.
\newblock Signed networks in social media.
\newblock In {\em Proceedings of the 28th International Conference on {{Human}}
  Factors in Computing Systems - {{CHI}} '10}, page 1361, 2010.

\bibitem{huangPOLEPolarizedEmbedding2022}
Zexi Huang, Arlei Silva, and Ambuj Singh.
\newblock {{POLE}}: {{Polarized Embedding}} for {{Signed Networks}}.
\newblock In {\em Proceedings of the {{Fifteenth ACM International Conference}}
  on {{Web Search}} and {{Data Mining}}}, pages 390--400, 2022.

\bibitem{garimellaLongTermAnalysisPolarization2017}
Venkata Rama~Kiran Garimella and Ingmar Weber.
\newblock A {{Long-Term Analysis}} of {{Polarization}} on {{Twitter}}.
\newblock {\em ICWSM}, 11(1):528--531, 2017.

\bibitem{gillaniMeMyEcho2018}
Nabeel Gillani, Ann Yuan, Martin Saveski, Soroush Vosoughi, and Deb Roy.
\newblock Me, {{My Echo Chamber}}, and {{I}}: {{Introspection}} on {{Social
  Media Polarization}}.
\newblock In {\em Proceedings of the 2018 {{World Wide Web Conference}} on
  {{World Wide Web}} - {{WWW}} '18}, pages 823--831, 2018.

\bibitem{tuckerSocialMediaPolitical2018}
Joshua~A. Tucker, Andrew Guess, Pablo Barbera, Cristian Vaccari, Alexandra
  Siegel, Sergey Sanovich, Denis Stukal, and Brendan Nyhan.
\newblock Social {{Media}}, {{Political Polarization}}, and {{Political
  Disinformation}}: {{A Review}} of the {{Scientific Literature}}, 2018.

\bibitem{grossDenseNetworkMotifs2023}
Bnaya Gross, Shlomo Havlin, and Baruch Barzel.
\newblock Dense network motifs enhance dynamical stability, 2023.

\bibitem{miloNetworkMotifsSimple2002}
R.~Milo, S.~{Shen-Orr}, S.~Itzkovitz, N.~Kashtan, D.~Chklovskii, and U.~Alon.
\newblock Network {{Motifs}}: {{Simple Building Blocks}} of {{Complex
  Networks}}.
\newblock {\em Science}, 298(5594):824--827, 2002.

\bibitem{ahmedGraphletDecompositionFramework2017}
Nesreen~K. Ahmed, Jennifer Neville, Ryan~A. Rossi, Nick~G. Duffield, and
  Theodore~L. Willke.
\newblock Graphlet decomposition: Framework, algorithms, and applications.
\newblock {\em Knowl Inf Syst}, 50(3):689--722, 2017.

\bibitem{przuljBiologicalNetworkComparison2007}
N.~Przulj.
\newblock Biological network comparison using graphlet degree distribution.
\newblock {\em Bioinformatics}, 23(2):e177--e183, 2007.

\bibitem{shen-orrNetworkMotifsTranscriptional2002}
Shai~S. {Shen-Orr}, Ron Milo, Shmoolik Mangan, and Uri Alon.
\newblock Network motifs in the transcriptional regulation network of
  {{Escherichia}} coli.
\newblock {\em Nat Genet}, 31(1):64--68, 2002.

\bibitem{tranCountingMotifsHuman2013}
Ngoc~Hieu Tran, Kwok~Pui Choi, and Louxin Zhang.
\newblock Counting motifs in the human interactome.
\newblock {\em Nat Commun}, 4(1):2241, 2013.

\bibitem{liInhomogeneousHypergraphClustering2017}
Pan Li and Olgica Milenkovic.
\newblock Inhomogeneous hypergraph clustering with applications.
\newblock In {\em Proceedings of the 31st {{International Conference}} on
  {{Neural Information Processing Systems}}}, {{NIPS}}'17, pages 2305--2315,
  2017.

\bibitem{chenNeMoFinderDissectingGenomewide2006}
Jin Chen, Wynne Hsu, Mong~Li Lee, and See-Kiong Ng.
\newblock {{NeMoFinder}}: Dissecting genome-wide protein-protein interactions
  with meso-scale network motifs.
\newblock In {\em Proceedings of the 12th {{ACM SIGKDD}} International
  Conference on {{Knowledge}} Discovery and Data Mining}, {{KDD}} '06, pages
  106--115, 2006.

\bibitem{spornsMotifsBrainNetworks2004}
Olaf Sporns and Rolf K{\"o}tter.
\newblock Motifs in {{Brain Networks}}.
\newblock {\em PLOS Biology}, 2(11):e369, 2004.

\bibitem{cartwrightStructuralBalanceGeneralization1956}
Dorwin Cartwright and Frank Harary.
\newblock Structural balance: A generalization of {{Heider}}'s theory.
\newblock {\em Psychological Review}, 63(5):277--293, 1956.

\bibitem{situngkirSocialBalanceTheory2004}
Hokky Situngkir and Deni Khanafiah.
\newblock Social {{Balance Theory}}: {{Revisiting Heider}}'s {{Balance Theory}}
  for many agents.
\newblock {\em Industrial Organization}, 2004.

\bibitem{szellMultirelationalOrganizationLargescale2010}
Michael Szell, Renaud Lambiotte, and Stefan Thurner.
\newblock Multirelational organization of large-scale social networks in an
  online world.
\newblock {\em Proceedings of the National Academy of Sciences},
  107(31):13636--13641, 2010.

\bibitem{raoMarkovChainMonte1996}
A.~Ramachandra Rao, Rabindranath Jana, and Suraj Bandyopadhyay.
\newblock A {{Markov Chain Monte Carlo Method}} for {{Generating Random}} (0,
  1)-{{Matrices}} with {{Given Marginals}}.
\newblock {\em Sankhy\=a: The Indian Journal of Statistics, Series A
  (1961-2002)}, 58(2):225--242, 1996.

\bibitem{singhMeasuringBalanceSigned2017}
Ranveer Singh and Bibhas Adhikari.
\newblock Measuring the balance of signed networks and its application to sign
  prediction.
\newblock {\em J. Stat. Mech.}, 2017(6):063302, 2017.

\bibitem{fengTestingBalanceSocial2022}
Derek Feng, Randolf Altmeyer, Derek Stafford, Nicholas~A. Christakis, and
  Harrison~H. Zhou.
\newblock Testing for {{Balance}} in {{Social Networks}}.
\newblock {\em Journal of the American Statistical Association},
  117(537):156--174, 2022.

\bibitem{kumarEdgeWeightPrediction2016}
Srijan Kumar, Francesca Spezzano, V.~S. Subrahmanian, and Christos Faloutsos.
\newblock Edge {{Weight Prediction}} in {{Weighted Signed Networks}}.
\newblock In {\em 2016 {{IEEE}} 16th {{International Conference}} on {{Data
  Mining}} ({{ICDM}})}, pages 221--230, 2016.

\bibitem{liConstructingRefinedNull2021}
Ai-Wen Li.
\newblock Constructing refined null models for statistical analysis of signed
  networks.
\newblock {\em Chinese Phys. B}, 30, 2021.

\bibitem{kirkleyBalanceSignedNetworks2019}
Alec Kirkley, George~T Cantwell, and M~E~J Newman.
\newblock Balance in signed networks.
\newblock {\em Physical Review E}, 99(1):11, 2019.

\bibitem{davisClusteringStructuralBalance1967}
James~A. Davis.
\newblock Clustering and {{Structural Balance}} in {{Graphs}}.
\newblock In {\em Social {{Networks}}}, pages 27--33. 1967.

\bibitem{esmailianMesoscopicAnalysisOnline2014}
Pouya Esmailian, Seyed~Ebrahim Abtahi, and Mahdi Jalili.
\newblock Mesoscopic analysis of online social networks: {{The}} role of
  negative ties.
\newblock {\em Phys. Rev. E}, 90(4):042817, 2014.

\bibitem{hararyMeasurementStructuralBalance1959}
Frank Harary.
\newblock On the measurement of structural balance.
\newblock {\em Behavioral Science}, 4(4):316--323, 1959.

\bibitem{bhatDensificationStructuralTransitions2016}
U.~Bhat, P.~L. Krapivsky, R.~Lambiotte, and S.~Redner.
\newblock Densification and structural transitions in networks that grow by
  node copying.
\newblock {\em Phys. Rev. E}, 94(6):062302, 2016.

\bibitem{cramerLogicalApproachRestricting2015}
Marcos Cramer, Jun Pang, and Yang Zhang.
\newblock A {{Logical Approach}} to {{Restricting Access}} in {{Online Social
  Networks}}.
\newblock In {\em Proceedings of the 20th {{ACM Symposium}} on {{Access Control
  Models}} and {{Technologies}}}, pages 75--86, 2015.

\bibitem{facchettiComputingGlobalStructural2011}
Giuseppe Facchetti, Giovanni Iacono, and Claudio Altafini.
\newblock Computing global structural balance in large-scale signed social
  networks.
\newblock {\em Proceedings of the National Academy of Sciences},
  108(52):20953--20958, 2011.

\bibitem{estradaWalkbasedMeasureBalance2014}
Ernesto Estrada and Michele Benzi.
\newblock Walk-based measure of balance in signed networks: {{Detecting}} lack
  of balance in social networks.
\newblock {\em Phys. Rev. E}, 90(4):042802, 2014.

\bibitem{marvelEnergyLandscapeSocial2009}
Seth~A. Marvel, Steven~H. Strogatz, and Jon~M. Kleinberg.
\newblock Energy {{Landscape}} of {{Social Balance}}.
\newblock {\em Phys. Rev. Lett.}, 103(19):198701, 2009.

\bibitem{harrisComputationalInferenceSynaptic2022}
Michael~R. Harris, Thomas~P. Wytock, and Istv{\'a}n~A. Kov{\'a}cs.
\newblock Computational {{Inference}} of {{Synaptic Polarities}} in {{Neuronal
  Networks}}.
\newblock {\em Advanced Science}, 9(16):2104906, 2022.

\bibitem{chiangExploitingLongerCycles2011}
Kai-Yang Chiang, Nagarajan Natarajan, Ambuj Tewari, and Inderjit~S. Dhillon.
\newblock Exploiting longer cycles for link prediction in signed networks.
\newblock In {\em Proceedings of the 20th {{ACM}} International Conference on
  {{Information}} and Knowledge Management}, pages 1157--1162, 2011.

\bibitem{hsiehLowRankModeling2012}
Cho-Jui Hsieh, Kai-Yang Chiang, and Inderjit~S. Dhillon.
\newblock Low rank modeling of signed networks.
\newblock In {\em Proceedings of the 18th {{ACM SIGKDD}} International
  Conference on {{Knowledge}} Discovery and Data Mining}, pages 507--515, 2012.

\bibitem{wangOnlineMatrixCompletion2017}
Jing Wang, Jie Shen, Ping Li, and Huan Xu.
\newblock Online {{Matrix Completion}} for {{Signed Link Prediction}}.
\newblock In {\em Proceedings of the {{Tenth ACM International Conference}} on
  {{Web Search}} and {{Data Mining}}}, pages 475--484, 2017.

\bibitem{yuanSNESignedNetwork2017}
Shuhan Yuan, Xintao Wu, and Yang Xiang.
\newblock {{SNE}}: {{Signed Network Embedding}}.
\newblock In Jinho Kim, Kyuseok Shim, Longbing Cao, Jae-Gil Lee, Xuemin Lin,
  and Yang-Sae Moon, editors, {\em Advances in {{Knowledge Discovery}} and
  {{Data Mining}}}, volume 10235, pages 183--195. 2017.

\bibitem{wangSignedNetworkEmbedding2017}
Suhang Wang, Jiliang Tang, Charu Aggarwal, Yi~Chang, and Huan Liu.
\newblock Signed network embedding in social media: 17th {{SIAM International
  Conference}} on {{Data Mining}}, {{SDM}} 2017.
\newblock {\em Proceedings of the 17th SIAM International Conference on Data
  Mining, SDM 2017}, pages 327--335, 2017.

\bibitem{islamSIGNetScalableEmbeddings2018}
Mohammad~Raihanul Islam, B.~Aditya~Prakash, and Naren Ramakrishnan.
\newblock {{SIGNet}}: {{Scalable Embeddings}} for {{Signed Networks}}.
\newblock In Dinh Phung, Vincent~S. Tseng, Geoffrey~I. Webb, Bao Ho, Mohadeseh
  Ganji, and Lida Rashidi, editors, {\em Advances in {{Knowledge Discovery}}
  and {{Data Mining}}}, volume 10938, pages 157--169. 2018.

\bibitem{chenBridgeEnhancedSigned2018}
Yiqi Chen, Tieyun Qian, Huan Liu, and Ke~Sun.
\newblock "{{Bridge}}": {{Enhanced Signed Directed Network Embedding}}.
\newblock In {\em Proceedings of the 27th {{ACM International Conference}} on
  {{Information}} and {{Knowledge Management}}}, pages 773--782, 2018.

\bibitem{javariROSERolebasedSigned2020}
Amin Javari, Tyler Derr, Pouya Esmailian, Jiliang Tang, and Kevin Chen-Chuan
  Chang.
\newblock {{ROSE}}: {{Role-based Signed Network Embedding}}.
\newblock In {\em Proceedings of {{The Web Conference}} 2020}, pages
  2782--2788, 2020.

\bibitem{yaoHowNetworkProperties2019}
Qing Yao, Tim~S. Evans, and Kim Christensen.
\newblock How the network properties of shareholders vary with investor type
  and country.
\newblock {\em PLOS ONE}, 14(8):e0220965, 2019.

\bibitem{kovacsUncoveringGeneticBlueprint2020}
Istv{\'a}n~A. Kov{\'a}cs, D{\'a}niel~L. Barab{\'a}si, and Albert-L{\'a}szl{\'o}
  Barab{\'a}si.
\newblock Uncovering the genetic blueprint of the {{{\emph{C}}}}{\emph{.
  elegans}} nervous system.
\newblock {\em Proc Natl Acad Sci USA}, 117(52):33570--33577, 2020.

\bibitem{chatterjeeImprovingGeneralizabilityProteinligand2023}
Ayan Chatterjee, Robin Walters, Zohair Shafi, Omair~Shafi Ahmed, Michael Sebek,
  Deisy Gysi, Rose Yu, Tina {Eliassi-Rad}, Albert-L{\'a}szl{\'o} Barab{\'a}si,
  and Giulia Menichetti.
\newblock Improving the generalizability of protein-ligand binding predictions
  with {{AI-Bind}}.
\newblock {\em Nat Commun}, 14(1):1989, 2023.

\end{thebibliography}

\end{document}